# On-chip tuning of the resonant wavelength in a high-Q microresonator integrated with a microheater

Jialei Tang,[1,2] Jintian Lin,[1] Jiangxin Song,[1,2] Zhiwei Fang, [1,3] Min Wang, [1,2] Yang Liao,[1] Lingling Qiao,[1] and Ya Cheng[1,*]

[1] *State Key Laboratory of High Field Laser Physics, Shanghai Institute of Optics and Fine Mechanics, Chinese Academy of Sciences, Shanghai 201800, China*

[2] *Graduate School of Chinese Academy of Sciences, Beijing 100039, China*

[3] *School of Physical Science and Technology, ShanghaiTech University, Shanghai 200031, China*

[*] *Electronic mail: ya.cheng@siom.ac.cn*

**Abstract:**

We report on fabrication of a microtoroid resonator of high-quality (high-Q) factor integrated with an on-chip microheater. Both the microresonator and microheater are fabricated using femtosecond laser three-dimensional (3D) micromachining. The microheater, which is located about 200 μm away from the microresonator, has a footprint size of 200 μm × 400 μm. Tuning of the resonant wavelength in the microresonator has been achieved by varying the voltage applied on the microheater. The drifting of the resonant wavelength shows a linear dependence on the square of the voltage applied on the microheater. We found that the response time of the microresonator is less than 10 secs which is significantly shorter than the time required for reaching a thermal equilibrium on conventional heating instruments such as an external electric heater.

Owing to their excellent capability to efficiently confine light in the periphery via total internal reflection, whispering-gallery-mode (WGM) optical microresonators have been playing increasingly important roles in applications such as microlasers [1], nonlinear optics [2], chemical and biological sensing [3,4], and cavity quantum electrodynamics (c-QED) [5]. In most of these applications, precise control of the resonant wavelength of WGM is highly desirable. The wavelength-tuning of WGM has been demonstrated by tactfully tuning the refractive index of the microresonator, which leads to a change in the optical path for the light travelling in the microresonator. Tuning of the refractive index is readily achievable by use of thermal-optical or electro-optical effects [6, 7]. In addition, the WGMs are also sensitive to the geometric shape of microresonator, enabling tuning of the resonant wavelength by slight distortion of the shape of the microresonator using mechanical stretching or squeezing [8,9]. To facilitate miniaturization and ease of operation, there is a pressing need for development of a monolithic approach by which the thermal or electric function can be integrated into the microresonator chip.

Recently, femtosecond laser direct writing has been proved to be a powerful approach for fabricating various types of three-dimensional (3D) micro-structures such as optical waveguides, microfluidics, micromechanics, and microelectrodes [10-14]. This technique therefore provides an ideal solution for integrating multifunctional microcomponents in a single chip, such as a microresonator integrated with a microheater which allows rapid tuning of the resonant wavelength. Specifically, 3D WGM microresonators of Q-factors on the order of $10^6$ have been fabricated in both fused silica and Nd:glass using femtosecond laser direct writing as reported in Refs. [15,16]. Meanwhile, selective metallization has also been achieved in transparent substrates such as glass and crystal based on femtosecond laser direct writing, which allows for the formation of microelectrodes and on-chip microheaters [14,17,18]. Nevertheless, these techniques have not been synergistically employed for achieving on-chip tuning of the resonant wavelength of a microresonator. In this Letter, we report on fabrication of a microresonator integrated with a microheater on a single fused silica chip by femtosecond laser direct writing. We demonstrate on-chip tuning of resonant wavelength of the microresonator by varying the electric voltage applied on the microheater. Since the microheater, whose footprint size is 200 μm × 400 μm, is only 200 μm away from the microresonator, the response time of the device is less than 10 secs which is significantly shorter than the time required for reaching a thermal equilibrium on conventional electric heaters.

The device is illustrated in Fig. 1(a), which contains the microheater and the microresonator. First, the microheater was fabricated by selective metallization on glass surfaces, which was realized using femtosecond laser direct writing followed by electroless plating. More details on the selective metallization can be found elsewhere [14,17]. The reason of fabricating the microheater in prior to the microresonator is for avoiding the contamination on the microresonator during the electroless plating. The fabrication process of microresonator is illustrated in Figs. 1(b-e), which mainly includes the fabrication of a freestanding microdisk supported by a thin pillar by femtosecond laser ablation of the glass substrate immersed in water, followed by $CO_2$ laser annealing for surface smoothing. More details about the fabrication of the microresonator can be found in Refs. [15,16].

In this work, commercially available fused silica glass substrates (UV grade fused silica JGS1, both upper and bottom surface were polished to optical grade) with a thickness of 1 mm were used. A Ti: Sapphire laser system (Coherent Inc.) with a central wavelength of 800 nm, a pulse width of 50 fs, and a repetition frequency of 250 kHz was used as the light source for direct writing. The laser beam was focused onto the upper surface of glass by an objective lens, and the sample could be arbitrarily translated by a PC-controlled XYZ stage with a resolution of 1 μm. A circular aperture with a diameter of 5 mm was used to improve the beam quality. The energy of the laser pulse could be adjusted using a neutral density (ND) filter and the power was monitored with an energy meter. The laser beam was focused by a 20× objective with a numerical aperture (N.A.) of 0.45, and the scan speed and the average power of the laser were set at 1 mm/s and 230 mW, respectively. For fabricating the microheater, microgrooves with a width of either 50 μm or 6 μm, and a depth of 15 μm were patterned on the glass surface by femtosecond laser raster scan. The two kinds of microgrooves of 50 μm- and 6 μm-width were designed to form the electrodes and resistors, respectively, with the subsequent electroless plating (Fig .1(a)).

To fabricate the microtoroid resonator, we used water-assisted femtosecond laser ablation which can directly produce a microdisk supported by a thin pillar without any chemical etching [16]. The microdisk and the underlying pillar were fabricated by a layer-by-layer annular ablation with a nanopositioning stage of the lateral scanning step set to be 1 μm, and the scanning speed was set at 600 μm/s. A 100× objective with a N.A. of 0.8 was used to focus the beam to a ~1 μm-dia. focal spot, and the average femtosecond laser power measured before the objective was ~1 mW. The diameter of the disk was 110 μm and the diameter of the pillar was 40 μm. We then smoothed the surface of the microresonator with thermal reflow. A $CO_2$ laser (Synrad Firestar V30) was employed for the surface reflow, which operated at a duty ratio of 5% and the surface reflow process lasted for ~4 secs.

The Q-factor of the microresonator was measured with the fiber taper coupling method [19]. A swept-wavelength tunable external-cavity New Focus diode Laser (Model: 6528-LN) and a dBm Optics swept spectrometer (Model: 4650) were used to measure the transmission spectra from the fiber taper with a resolution of 0.1 pm. The fiber taper was fabricated by heating and stretching a section of bare optical fiber (SMF-28, Corning) until reaching a minimum waist diameter of ~1 μm. The fabricated microresonator was fixed on a nano-positioning stage with a spatial resolution of 50-nm in XYZ directions so that the critical coupling could be realized by carefully adjusting the relative positions between the microresonator and the fiber taper. We used dual CCD cameras to simultaneously image microtoroid resonator and fiber taper from the side and the top view.

Fig. 2(a) presents the optical micrograph (close-up view) of the central area of the microheater, highlighting the smooth surface and homogeneous width of the copper lines formed by femtosecond laser selective metallization. Our previous investigations have shown that the metal structures are buried in the substrate which ensures a good adhension [18]. The total resistance (R) of microheater was measured to be R=9.6 Ω. At an electric voltage of 1 V, the power of the microheater is calculated to be ~0.1 W. The microheater was connected to the power supplier by two copper wires which were bonded on the two ends of the microheater using conductive silver adhesive. This procedure was arranged after the microresonator was fabricated, because the $CO_2$ laser irradiation employed for surface reflow on the microresonator will cause damage to the silver adhesive. Figs. 2(b)-2(c) show the side and top view scanning electron micrograph (SEM) images of the fabricated mocroresonator, respectively. The Q-factor of microresonator was measured to be $\sim 1.2 \times 10^6$ under the critical coupling condition, as evidenced by the Lorentz fitting in Fig. 2(d).

After the highest Q factor was achieved, we maintained all the conditions of the experimental system but only allowed the electric voltage of the power supplier to vary. In such a manner, we were able to examine the dependence of resonant wavelength on the electric voltage applied on the microheater. The temperature increase resulted in a red-shift of the resonant wavelengths. Almost immediately after applying the electric voltage on the microheater, the spectra started to drift on the spectrometer. After the electric voltage was changed, it took ~10 secs for the system to reach a stabilized temperature. Afterwards, the transmission spectra became stable on the spectrometer.

The transmission spectra of microresonator in Fig. 3(a) were recorded at different electric voltages of 0 V, 1 V, 2 V and 3 V. The central wavelength of the WGM shifts from 1561.97 nm (the black solid line) to 1562.15 nm (the pink dash dot line) due to the temperature change in the microtoroid resonator as the power of microheater increased from 0 W to 0.9 W. Fig. 3(b) shows that the drifting of the resonant wavelength is linearly dependent on the $V^2$ of the microheater, because the Joule heating power of the microheater is proportional to $V^2$. The slope of the fitting line indicates that the tuning rate of the microheater is 1.8 GHz/$V^2$.

We also examined the response time and the stability of the wavelength tuning realized by the integrated microheater. For this purpose, we quickly raised the electric voltage to 2 V and then maintained this voltage for the subsequent measurements. We recorded the transmission spectrum of microresonator at each 5-second interval. Fig. 4(a) shows the result of the spectra recorded at different times. The solid black line shows the spectrum measured at the electric voltage of 0 V (i.e., before applying any voltage on the microheater). The dashed red line shows the spectrum measured 5 secs after the voltage was applied on the microheater. The resonant wavelength has been shifted by ~100 pm due to the heating of microresonator in the first 5 secs, as indicated by the red arrow. The spectra became stabilized after the voltage was applied for 10 secs. This is evidenced by the spectra measured 10 secs (the dotted blue curve) and 15 secs (the dash doted green curve) after the voltage was switched to 2 V, which shows little difference from each other. We would like to point out that to realize on-chip tuning of resonant wavelength with an external heater, a significantly longer time is required to reach the thermal equilibrium (e.g., a few minutes as reported in [20]). The microheater directly integrated near the microresonator with femtosecond laser direct writing not only leads to a compact device but also a faster response by efficient local heating in a confined area.

In conclusion, we have demonstrated fabrication of a high-Q microtoroid resonator integrated with a microheater on a single fused silica chip by femtosecond laser micromachining. On-chip tuning of the WGM microresonator has been accomplished with the integrated microheater. The response time of our device is on the order of ~10 secs, which significantly surpass the performance of conventional external heaters. Our technique can benefit various applications including tunable laser sources, optical filters, cavity quantum electrodynamics, etc.

This research is financially supported by National Basic Research Program of China (No. 2014CB921300), National Natural Science Foundation of China (Nos. 61275205, 61108015, and 11104294).

**References:**

[1] S. L. McCall, A. F. J. Levi, R. E. Slusher, S. J. Pearton, and R. A. Logan, Appl. Phys. Lett. **60**, 289 (1992).

[2] S. M. Spillane, T. J. Kippenberg, and K. J. Vahala, Nature **415**, 621 (2002).

[3] F. Vollmer and S. Arnold, Nature Methods **5**, 591 (2008).

[4] A. M. Armani, R. P. Kulkarni, S. E. Fraser, R. C. Flagan, and K. J. Vahala, Science **317**, 783 (2007).

[5] D. J. Alton, N. P. Stern, Takao, H. Lee, E. Ostby, K. J. Vahala, and H. J. Kimble, Nature Phys. **7**, 159 (2011).

[6] J. M. Ward and S. N. Chormaic, Appl. Phys. B **100**, 847 (2010).

[7] J. P. Rezac and A. T. Rosenberger, Opt. Express **8**, 605 (2001).

[8] K. N. Dinyari, R. J. Barbour, D. A. Golter, and H. Wang, Opt. Express **19**, 17966 (2011).

[9] D. Armani, B. Min, A. Martin, and K. J. Vahala, Appl. Phys. Lett. **85,** 5439 (2004).

[10] R. R. Gattass and E. Mazur, Nat. Photonics **2**, 219 (2008).

[11] K. Sugioka and Y. Cheng, Light: Sci. Appl. **3**, e149 (2014).

[12] R. Osellame, H.J.W.M. Hoekstra, G. Cerullo, and M. Pollnau, Laser Photon. Rev. **5**, 442 (2011).

[13] F. Chen and J. R. Vázquez de Aldana, Laser Photon. Rev. **8**, 251 (2014).

[14] Y. Liao, J. Xu, Y. Cheng, Z. H. Zhou, F. He, H. Y. Sun, J. Song, X. S. Wang, Z. Z. Xu, K. Sugioka, and K. Midorikawa, Opt. Lett. **33**, 2281 (2008).

[15] J. Lin, S. Yu, Y. Ma, W. Fang, F. He, L. Qiao, L. Tong, Y. Cheng, and Z. Xu, Opt. Express **20**, 10212 (2012).

[16] J. Lin, S. Yu, J. Song, B. Zeng, F. He, H. Xu, K. Sugioka, W. Fang, and Y. Cheng, Opt. Lett. **38**, 1458 (2013).

[17] J. Xu, Y. Liao, H. Zeng, Z. Zhou, H. Sun, J. Song, X. Wang, Y. Cheng, Z. Xu, K. Sugioka, and K. Midorikawa, Opt. Express **15**, 12743 (2007).

[18] Y. Liao, J. Xu, H. Sun, J. Song, X. Wang, and Y. Cheng, Appl. Surf. Sci. **254**, 7018 (2008).

[19] A. Serpengüzel, S. Arnold, and G. Griffel, Opt. Lett. **20**, 654 (1995).

[20] A. Chiba, H. Fujiwara, J. Hotta, S. Takeuchi, and K. Sasaki, Jpn. J. Appl. Phys. **43**, 6138 (2004).

**Figure Captions:**

Fig. 1: (a) Illustration of the layout of the integrated chip. (b) Fabrication of the microdisk structure by water assisted femtosecond laser direct ablation. (c) Illustration of the fabricated microdisk. (d) Selective reflow of the silica microdisk by $CO_2$ laser irradiation. (e) Illustration of the microtoroid resonator.

Fig. 2: (a) The optical micrograph of the central part of microheater composed of metal lines. (b) Side view and (c) top view of the scanning electron micrograph (SEM) image of the microresonator. (d) Lorentz fitting (red curve) of the transmission spectrum near a resonant wavelength, showing a Q-factor of $1.2 \times 10^6$.

Fig. 3: (a) Normalized transmission spectra of the microresonator at different electric voltages applied on the microhearter. (b) The drift of resonant wavelength as a function of square of the voltage, which can be reasonably fitted with a linear relationship (red cure).

Fig. 4: (a) Normalized transmission spectra of the microresonator measured before (black curve) and 5s (red dashed), 10s (blue dotted), and 15s (green dash-dotted) after an electric voltage of 2 V was applied on the microhearter. (b) Close-up view of the transmission spectra in the spectral range indicated by the red box in (a). The red arrows in (a) and (b) indicate the shift of resonant wavelength induced by the heating of the resonator.

Fig. 1

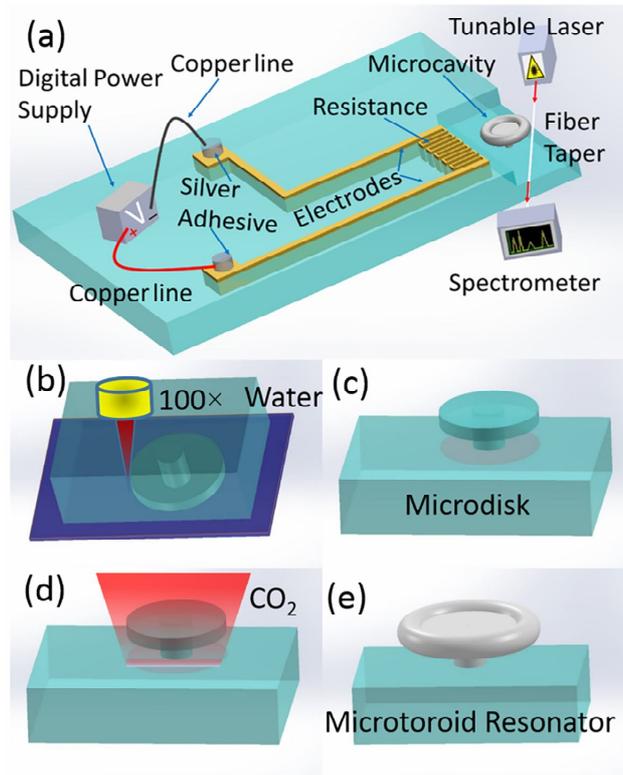

Fig. 2

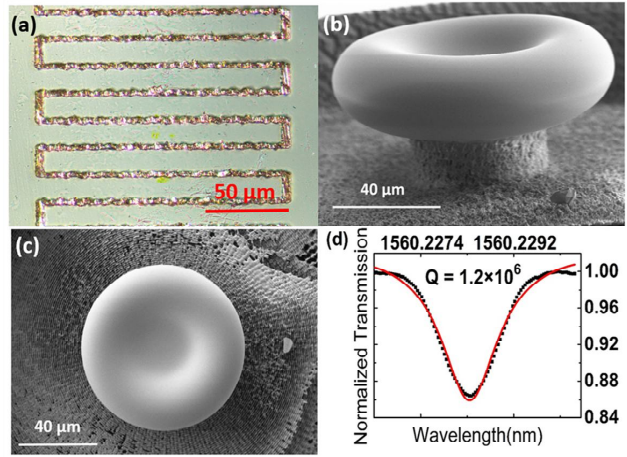

Fig. 3

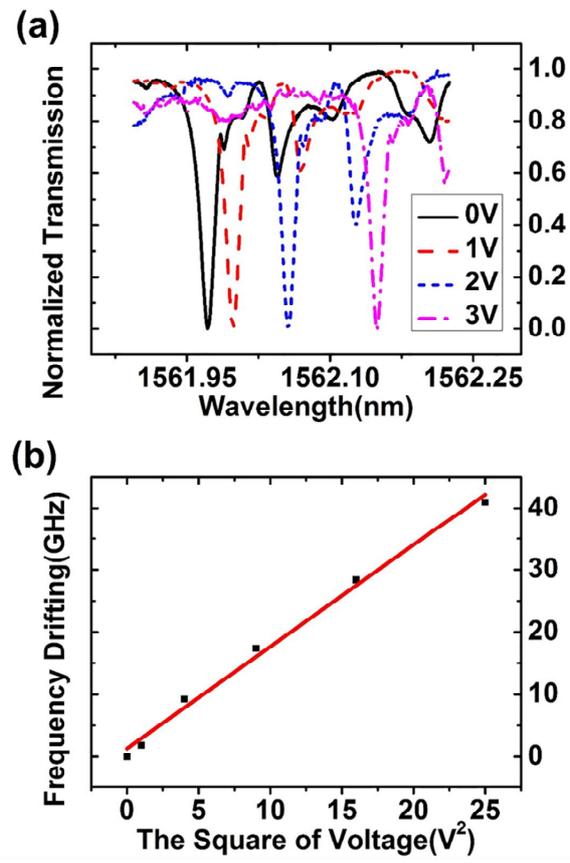

Fig. 4

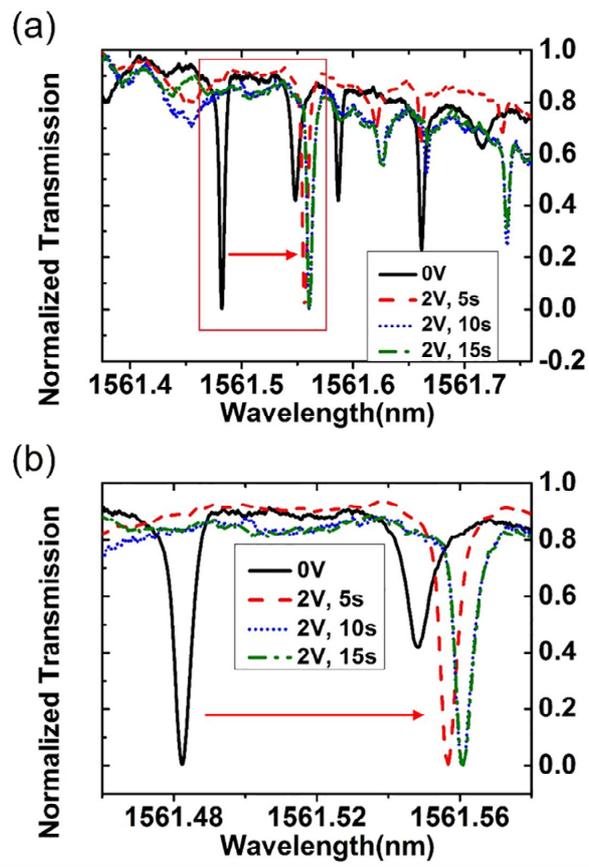